\documentclass[11pt]{article}
\usepackage{moriond,epsfig}

\def\be{\begin{equation}}
\def\ee{\end{equation}}
\def\bea{\begin{eqnarray}}
\def\eea{\end{eqnarray}}

\usepackage[switch,pagewise]{lineno}
\usepackage{wrapfig}
\usepackage[hyphens]{url}
\usepackage{subfig}

\begin{document}
\vspace*{4cm}
\title{The Shadow of the Moon in IceCube}

\author{ L. Gladstone, for the IceCube Collaboration
\footnote{For a complete author list, see http://www.icecube.wisc.edu/collaboration/authorlists/2010/4.html}
}

\address{University of Wisconsin-Madison, 1150 University Avenue, Madison, WI 53706, USA}

\maketitle\abstracts{
IceCube is the world's largest neutrino telescope, recently
completed at the South Pole. As a proof of pointing accuracy, we look
for the image of the Moon as a deficit in down-going cosmic ray muons,
using techniques similar to those used in IceCube's astronomical
point-source searches.}

\section{Introduction}\label{sec:Intro}

\begin{wrapfigure}{r}{8cm}
	\centering
		\includegraphics[trim=10 0 10 10,clip,scale=0.33]{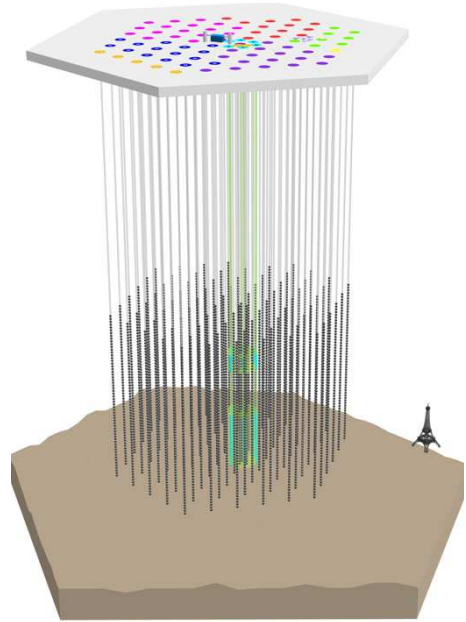}
	\caption{A schematic view of the IceCube detector, with the Eiffel Tower added for scale. There are 86 strings of detector modules deployed within the glacier; where each string connects to the surface, there is a dot. The color of the dot represents the year which that string was deployed.}
	\label{fig:detector}
\end{wrapfigure}

One of the main goals of the the IceCube detector\cite{detector} at the South Pole is to look for astrophysical point sources of neutrinos: essentially, IceCube is a telescope using neutrinos instead of light. 
Other telescopes can calibrate their signals with known standard candles (the Crab nebula is a traditional standard candle in gamma ray astronomy, for example). 
In the absence of known high-energy neutrino source, IceCube can use the deficit of cosmic ray muons from the direction of the Moon for calibration. This ``Moon Shadow'' is valuable because of its well-known position. 
While a muon calibration is not as directly applicable to neutrino astronomy as a Crab gamma calibration would be to gamma astronomy, it does provide valuable information about the pointing accuracy and resolution of the detector.

The 86-string IceCube detector was built modularly, with strings deployed during 7 consecutive austral summers. 
The changing detector size thus creates an annual discreteness in the data.
A Moon shadow analysis was developed for several of the detector setups; this work focuses on a 40-string setup analysis~\cite{ICRC} and a 59-string setup analysis~\cite{Jan}$^,$\cite{Hugo}.

\section{Data Sample}\label{sec:DataSample}
Because of bandwidth restrictions on the satellite transporting data from the South Pole to the North for analysis, a subset of available IceCube data is used. For these analyses, the data were collected in the following way: tracks were reconstructed quickly, and their direction of origin was compared to the current position of the Moon. If an event came from a position within $40^{\circ}$ in azimuth or $10^{\circ}$ in zenith, it was sent north. These data were collected only when the Moon was $15^{\circ}$ or more above the horizon at the South Pole, which neatly splits the data into lunar months. This sample was used for both a Moon measurement and an off-source background estimate.

The estimated angular resolution of the reconstructions used here is of order $1^{\circ}$, similar to the $0.5^{\circ}$-diameter Moon, so for this analysis the Moon was considered point-like. 

\section{Binned 40-string analysis}
\begin{figure}[htbp]
	\centering
		\includegraphics[scale=0.25]{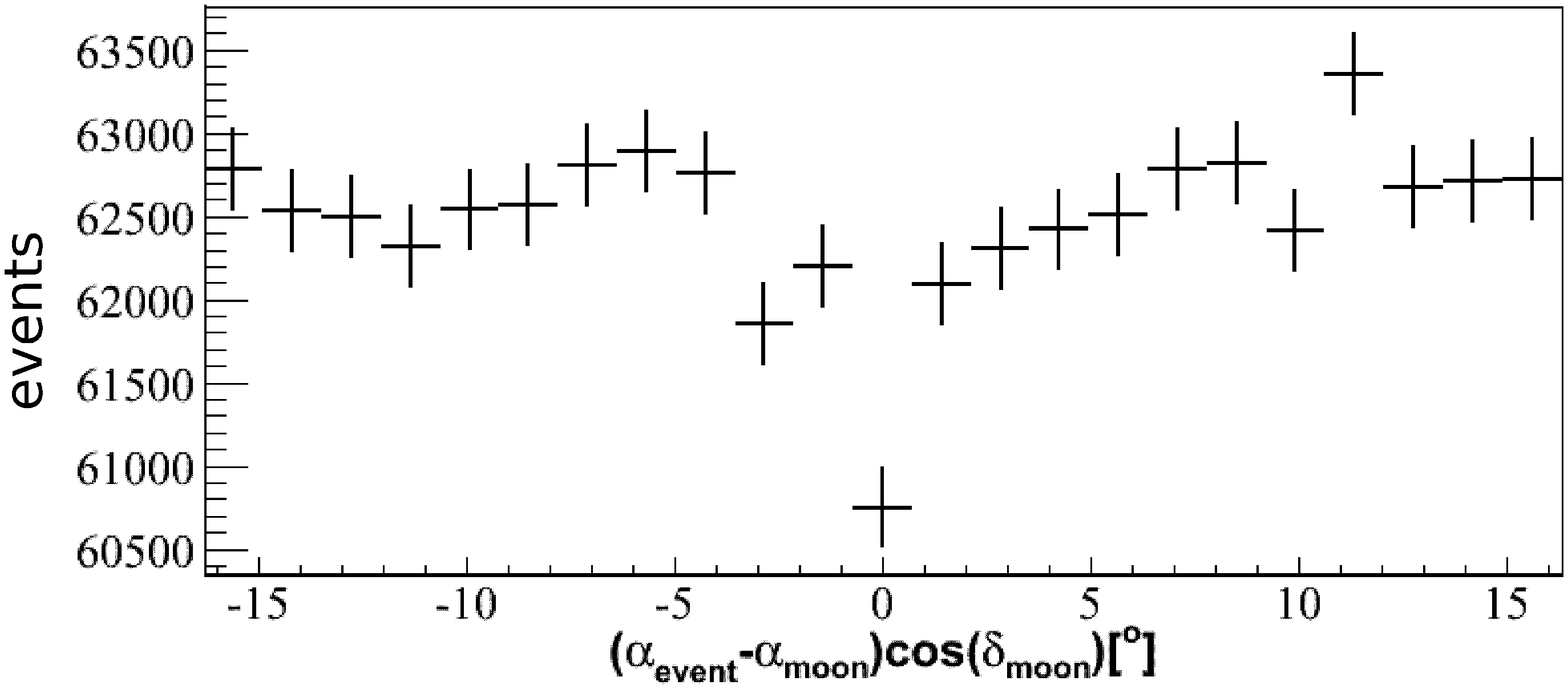}
	\caption{PRELIMINARY: Events in a $1.25^{\circ}$ zenith band around the Moon, using the 40-string detector setup. A deficit from the direction of the Moon can be clearly seen at 0. }
	\label{fig:binned1D}
\end{figure}

One analysis of the Moon Shadow was performed on the data set from the 40-string detector setup. This dataset contained 13 lunar months. 
Cuts were applied to the data sample to optimize the expected signal
(balancing passing rate with the expected improvement to the point spread function). 
Using simulation, the search bin size was optimized, and a band $1.25^{\circ}$ tall at constant zenith with respect to the Moon was used. Figure~\ref{fig:binned1D} shows the number of events in this zenith band, using the same optimized bin size of $1.25^{\circ}$ in azimuth as in zenith. Taking the mean of all the bins excluding the central 4 as a comparison, and using simple statistical $\sqrt{N}$ errors, we see a deficit of $7.6\sigma$ in the central bin at the position of the Moon.

\section{Likelihood 59-string analysis}
A subsequent analysis\cite{Jan} used data from the 59-string setup of the detector. The approach for this analysis was similar to the likelihood approach taken for the IceCube point source searches: an expected signal shape and background shape were developed, and then a likelihood was maximized at every point in the sky, allowing the number of signal events to vary. The likelihood formula used is: 
\[
L(\vec{x_s},n_s) = 
\sum_i^N{\log\left( \frac{n_s}{N}S_i + (1 - \frac{n_s}{N})B_i \right)}
\]
where $\vec{x_s}$ is the position being considered (relative to the Moon), $n_s$ is the number of signal events, $N$ is the total number of events, $S_i$ is the expected signal shape, and $B_i$ is the expected background shape. Note that this has no explicit energy term; this a major difference between the IceCube Moon analysis and the IceCube point source searches\cite{PointSource}. For the Moon shadow, we expect the number of signal events to be negative, as the Moon produces a deficit. 

Each event's contribution to the signal shape was assumed to be gaussian, with a width given by the estimated error on the reconstructed position. 

The background shape was estimated using two off-source regions: to the right and left of the moon in azimuth, at the same zenith. For each region, the event rate was assumed constant in azimuth, and an 80 bin histogram (with interpolation between bins) was used to describe the zenith distribution.

\begin{figure}[htbp]
	\centering
		\subfloat[Fluctuations around the background model of an off-source region, defined in relation to the position of the Moon. The color axis is the best-fit number of total ``signal'' events given the response at that point.]{
			{\label{fig:background_histo}}
			\includegraphics[trim=0 0 -190 0,clip,width=0.46\columnwidth]{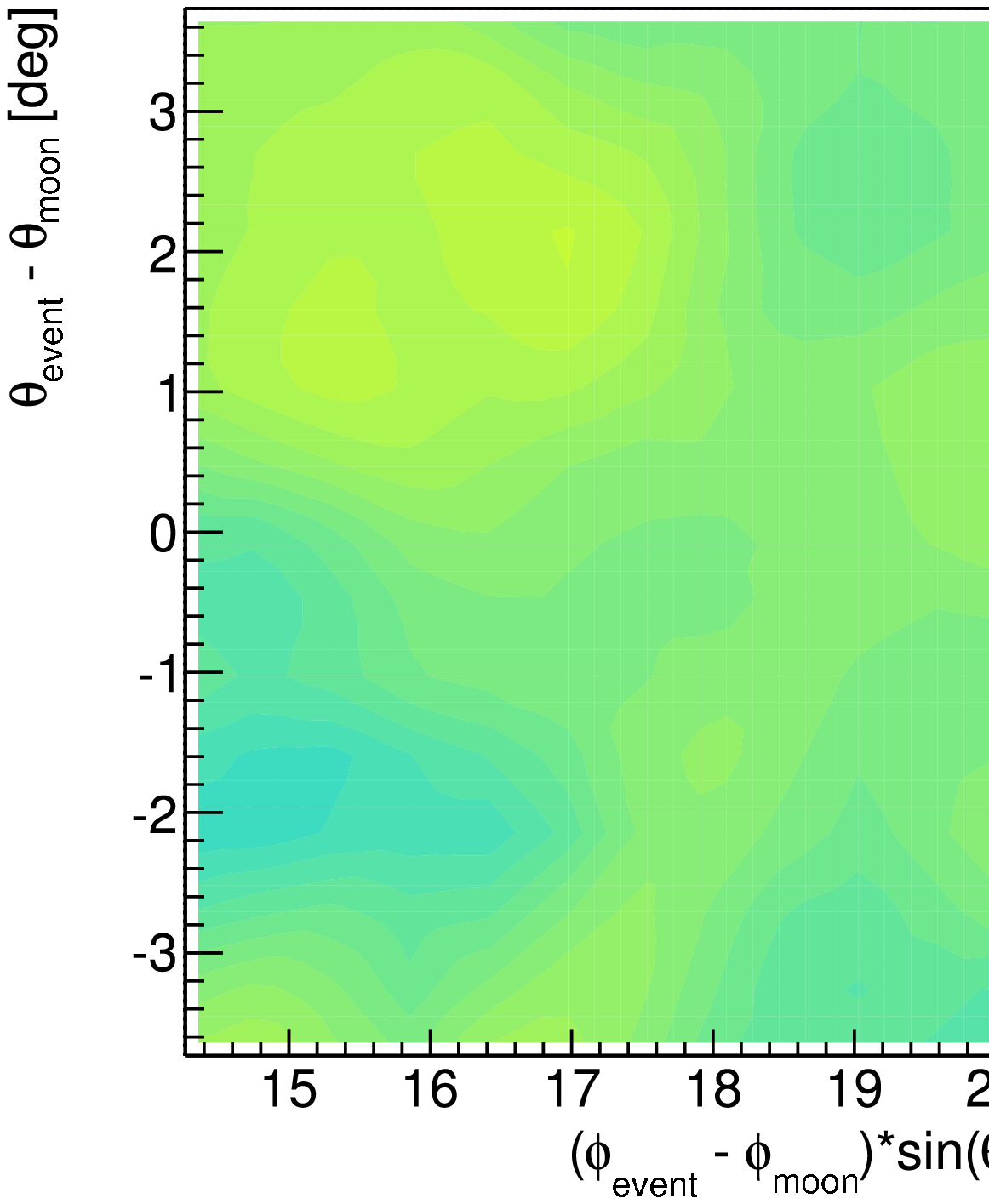}
		}
		\hspace{10pt}
		\subfloat[The distribution of bins from Fig.~\ref{fig:background_histo}, which can be fit to a gaussian curve, confirming that the background  is fluctuating randomly around 0.]{
			{\label{fig:pull}}
			\includegraphics[trim=0 -30 -180 -30,clip,width=0.46\columnwidth]{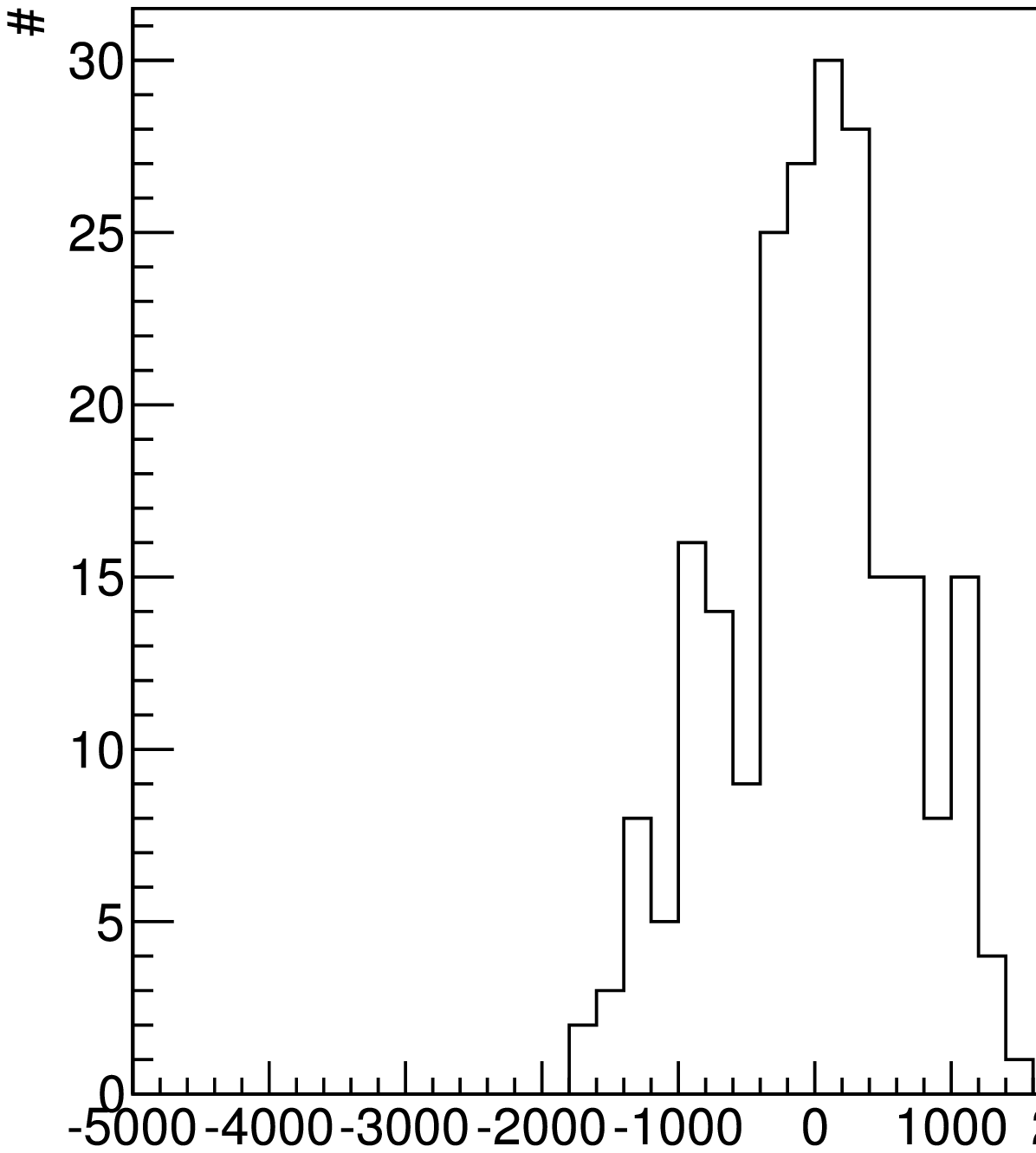}
		}
	\caption{PRELIMINARY: from~[3]}
	\label{fig:background}
\end{figure}

One can test the quality of this background model by assuming it as the background truth, and examining the size of fluctuations in the background region (calling the background region data ``signal'' for the purposes of this test).
For a perfect background model, this should result in only random fluctuations around zero. 
The result is shown in Figure~\ref{fig:background_histo}. 
The fluctuations of the background show up in the figure as indicated by the color axis. 
To test that these fluctuations are random, the value of each bin from Figure~\ref{fig:background_histo} is plotted in Figure~\ref{fig:pull}. 
The distribution of these values is consistent with a Gaussian fit centered at 0. 
The rms width of this distribution is about 680 signal events, which we take as the definition of $1\sigma$. 
A similar analysis was performed on the other background sample, resulting in a width of 560 events. 
As the two rms values were slightly different, the significance reported here should be taken only approximately. 
We consider the wider fluctuation value of 680, to be conservative. 


\begin{figure}[htbp]
	\centering
	\includegraphics[scale=0.5]{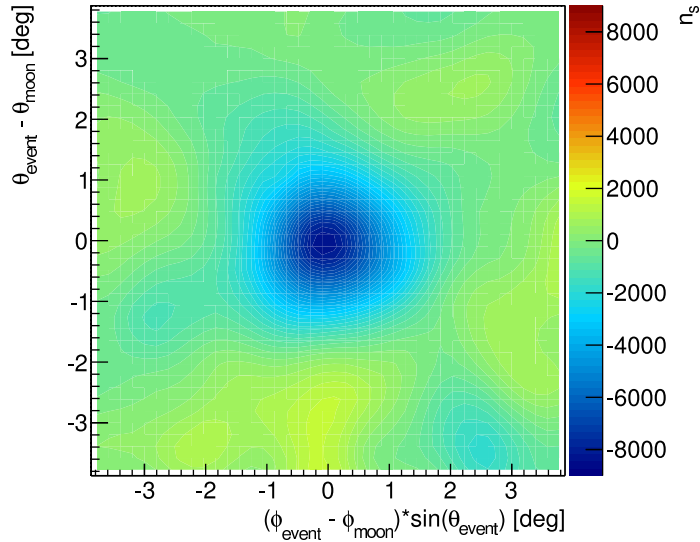}	
	\caption{PRELIMINARY: The Moon Shadow from the 59-string detector setup, using a likelihood analysis approach. The position is given relative to the Moon position, and the color represents the number of total shadowed events for each point, assuming the Moon is at that point. 	
	From~[3].}
	\label{fig:signal}
\end{figure}

This procedure is then applied to the signal region around the known Moon position: the binned and interpolated zenith model as a background description, and the sum of the observed data as the signal. The resulting plot is shown as Figure~\ref{fig:signal}. Each point represents the number of events shadowed if the Moon were at that point; the maximum of these is at the expected position of the Moon, with 8660 events shadowed. 
Taking $1\sigma  = 680$ events as discussed above, this is a $12.7\sigma$ observation. 

The expected number of shadowed events, based on the background rate and the size of the Moon, was $8192\pm91$. The observation of a 8660 event deficit at the central grid position is within $1\sigma$ of the expectation.

\section{Conclusions}
In each of two years of data during the construction of the IceCube detector, a shadowing effect was observed in cosmic rays from the direction of the Moon. In the 40-string setup, this deficit was observed with $7.6\sigma$ using a binned analysis. In the 59-string setup, this deficit was observed with $12.7\sigma$ using a likelihood analysis. The results confirms the pointing resolution of IceCube to within order $1^{\circ}$. Further studies of this shadowing effect are forthcoming.

\section*{Acknowledgments}
We gratefully acknowledge support from the National Science Foundation, and LG acknowledges that this research was made with Government support under and awarded by DoD, Air Force Office of Scientific Research, National Defense Science and Engineering Graduate (NDSEG) Fellowship, 32 CFR 168a.

\end{document}